# Laser cooling of a potassium-argon gas mixture using collisional redistribution of radiation

Anne Saß, Ulrich Vogl, and Martin Weitz

*Institut für Angewandte Physik der Universität Bonn, Wegelerstraße 8, D-53115 Bonn*

Abstract

We study laser cooling of atomic gases by collisional redistribution, a technique applicable to ultradense atomic ensembles at a pressure of a few hundred bar. Frequent collisions of an optically active atom with a buffer gas shift atoms into resonance with a far red detuned laser beam, while spontaneous decay occurs close to the unperturbed resonance frequency. In such an excitation cycle, a kinetic energy of the order of the thermal energy $k_B T$ is extracted from the sample. Here we report of recent experiments investigating the cooling of a potassium-argon gas mixture, which compared to an rubidium-argon mixture investigated in earlier experiments has a smaller fine structure of the optically active alkali atom. We observe a relative cooling of the potassium-argon gas mixture by 120 K.

## 1. Introduction

Light as a tool for cooling of matter was considered for the first time already in 1929 by Pringsheim [1]. In the last 25 years, Doppler cooling of dilute atomic gases has developed as a very widely used application of this concept [2, 3], and more recently anti-Stokes fluorescence cooling in multilevel systems has been investigated, allowing for an optical refrigeration of solids [4-6]. In this paper, we study the cooling of an ultradense alkali - noble gas mixture by collisional redistribution of radiation. Redistribution of atomic fluorescence is most widely known in the context of magneto-optic trapping of ultracold atoms, where the mechanism is a primary cause of trap loss processes [7]. In the long investigated field of room-temperature atomic collisions, redistribution of fluorescence is a natural consequence of line broadening effects from collisional aided excitation [8]. In theoretical works, Berman and Stenholm proposed laser cooling and heating of two-level systems based on the energy loss respectively during collisional aided excitation of atoms [9]. Experiments carried out with gases of moderate density have observed a heating for blue detuned radiation, but the cooling regime was never reached [10]. In a recent experiment, our group has demonstrated the relative cooling of a gas of rubidium atoms with 230 bar of argon atoms by 66 K by collisional redistribution of radiation [11]. The cooled gas has a density of more than ten orders of magnitude above the typical values in Doppler cooling experiments. We here report on experiments where a potassium-argon mixture was used for cooling. The potassium atom has a smaller D-lines fine structure in the electronically excited level than the rubidium atom, which equals 3 nm in the case of the potassium atom instead of the value of 14 nm for the rubidium case. The pressure broadened linewidth in the high pressure buffer gas system reaches a few nm (it approaches the thermal energy $k_B T$ in wavelength units), so that there is hope that the smaller fine structure splitting of the potassium atom makes the system closer to a two-level system, which in the future can facilitate the interpretation of results. We describe the current status of the experiment and give an overview of results obtained in the potassium - argon system. In the following, Section 2 describes the principle of the cooling method, Section 3 the used experimental setup and Section 4 contains the obtained results. Finally, Section 5 gives conclusions and an outlook.

## 2. Cooling principle

We begin by describing the principle of redistribution laser cooling of the dense gas system. Our experiment uses potassium atoms at thermal vapour pressure in 200 bar of argon buffer gas. Fig. 1 indicates the variation of the potassium 4s ground state and the 4p excited state respectively versus distance from an argon gas perturber in a binary collisional picture. Note that the potential curves are sketched schematically, a more realistic description can be found in [12]. The used cooling laser was detuned from the atomic resonance by a few nanometers to the red, i.e. 15 nm from the unperturbed potassium D1-line. Excitation becomes feasible when a noble-gas atom approaches one alkali atom and transiently shifts the D-lines into resonance. The 28 ns natural lifetime of the potassium 4p state is four magnitudes larger than the duration of a collisional process. The collisional rate in the high pressure buffer gas environment is $\approx 10^{11}$/s. We here assume quasi-elastic collisions, in a sense that no quenching effects occur, and a range of the potential much smaller than the interatomic spacing. Subsequent spontaneous decay occurs most likely at larger distance of atoms and perturbers, i.e. close to the unperturbed resonance frequency, see Fig. 1. During each excitation cycle, a kinetic energy of order of the thermal energy $k_B T$ is extracted from the dense atomic sample, which is of the same order of magnitude than the used laser detuning in energy units. Note that early experiments aiming towards a redistribution cooling in the moderate pressure regime were limited to detunings from resonance much below the possible energy exchange of $k_BT$, due to the smaller value of the pressure broadening [10]. In general, the here described simple picture neglects multiparticle collisions, which can be relevant at the used buffer gas pressures of a few 100 bar used in the present experiments [14].

## 3. Experimental Setup

The experiment takes place in a stainless-steel high pressure chamber with sapphire optical windows. Because of its lower vapour pressure at constant temperature with respect to the rubidium atom, in the here investigated potassium case the pressure cell has to be heated to higher temperatures then in the rubidium case reported earlier [13], to achieve a sufficient optical density for the absorption of the cooling laser beam. For the cooling experiments, the atomic ensemble (200 bar argon and thermal vapour pressure of potassium, $n_{Ar} \approx 10^{21}$ cm$^{-3}$, $n_P \approx 10^{16}$ cm$^{-3}$) was initially heated up to 440°C to achieve sufficient thermal potassium vapour pressure to be optically dense. To reduce quenching of excited state potassium atoms by collisions with impurities in the buffer gas, we use argon buffer gas of purity 6.0. Cooling radiation near the potassium D-lines is generated with a continous-wave Ti:Sapphire ring laser, whose output power is 3 W. The incident laser beam is guided through the cell volume through two sapphire optical windows with 18 mm diameter. The optical path length inside the cell is 10 mm. Figure 2 shows the experimental setups for the recording of fluorescence spectra (Fig. 2a) and temperature measurements inside the cell (Fig. 2b) respectively, as described in the following. A confocal setup is used to record spectra of fluorescence of the high pressure sample. To suppress a background of scattered light backscattered mostly from the cell windows from detection, both the incident beam and the outgoing fluorescence are spatially filtered with pinholes. The fluorescence signal is subsequently detected by an optical spectrum analyzer. For a measurement of the temperature change in the gas cell induced by the cooling beam, we employed thermal deflection spectroscopy [17, 18]. For this measurement of the cooling, the Ti:sapphire laser was tuned to a wavelength of 785 nm, far to the red of the unperturbed potassium D-lines (that is 766 nm and 770 nm, for the D2- and D1-lines respectively). Inside a telescope setup, an optical chopper is used to periodically block the cooling laser beam. Thermal deflection spectroscopy of the temperature change of the gas induced by the cooling beam near its beam path is applied by directing a non-resonant helium-neon laser beam collinear to the cooling laser path, with a variable lateral spatial offset. The temperature gradient induced by the cooling beam causes a spatial variation of the refractive index, resulting in prismatic deflection of the probe beam. Its angular deflection can be detected by a position sensitive photodiode placed behind the cell. Both the cooling laser beam and atomic fluorescence are blocked by a shortpass filter located in front of the photodiode. With a translation stage, the transverse position of the cooling laser beam can be varied with respect to the probe laser beam, so that the temperature profile induced by the cooling beam in the gas can be mapped out.

## 4. Results and Discussion

In the following, experimental results for the spectroscopy of potassium atoms in the high pressure buffer gas environment and that of temperature measurements based on thermal deflection spectroscopy measurements will be discussed.

**4.1. Atomic Fluorescence**

Using the described confocal setup, we recorded atomic fluorescence spectra from the dense buffer gas broadened sample. The spectra were measured under constant high-pressure buffer gas conditions for a variable incident laser detuning. Figure 3 presents two typical spectra with incident wavelengths of 765 nm and 780 nm respectively. The laser beam power was 630mW on the 3µm beam radius used for this measurement. For both measurements, a clear redistribution of the wavelength to the center of the D-lines is observed. Note that the D-lines center are pressure shifted to roughly 770nm and 773nm wavelength for the D2- and D1-lines respectively, i.e. by some 3-4 nm to the red compared to the position of the unperturbed case. The spectra also show a peak at the spectral position of the incident light, which is attributed to residual scattered light from the incident laser beam. This could be further reduced with improved spatial filtering (the present experiment due to low optical density of the potassium vapour used somewhat larger pinholes than required for a true mode-matched confocal filtering, given the low intensity of the backscattered light). For an incident radiation of 780 nm (see the dashed red line for the corresponding fluorescence spectrum) the redistribution towards the lines center is expected to extract energy from the sample, which can result in a laser cooling. On the other hand for an incident radiation of 765 nm (see the solid black line in the spectrum), which is blue detuned, we expect heating of the sample. In general, on the red wing of the potassium D-lines, a much higher fluorescence level with several smaller resonances is observed than on the blue wing (lower wavelengths). We attribute the resonances to satellite bands of potassium-argon quasi-molecules [15, 16] and the shape of the spectrum as being determined by the overlap of the corresponding quasi-molecular wavefunctions.

**4.2 Laser cooling of the gas**

To test for redistribution cooling of the dense gas, the spatial temperature profile induced by the cooling beam was monitored by thermal deflection spectroscopy. For this measurement, the lateral offset between the cooling and the probe beam was varied in steps of 50μm. Temperature variations induced by the cooling beam cause a local density change and a change of the refractive index n of the gas according to dn/dT≈-(n-1)/T, developed from the Lorentz-Lorenz relation [19], which incorpates that (n-1) is proportional to the gas density for n ≈ 1. This yields a prismatic deflection of the probe beam. The expected deflection angle after the passage of this beam through the cell of length L is:

$$\varphi = \frac{n-1}{T} \int_0^L \frac{dT(r,z)}{dr} dz,$$

where r is the transverse displacement between the cooling and probe beams. Assuming that the cooling rate is of the same shape as the Gaussian intensity profile of the incident cooling beam $I(r,z) = I_0 \exp(-\alpha z - 2r^2/w^2)$, where α=1/$l_{abs}$ denotes the absorption coefficient, for 1/α >> w, the heat transfer in the axial direction can be neglected. The deflection of the probe beam is determined by the transverse temperature gradient and the temperature distribution can be determined using:

$$\Delta T(r,z) = \frac{T}{(n-1)} \cdot \frac{\alpha e^{-\alpha z}}{1-e^{-\alpha L}} \int_r^\infty \varphi(r')dr',$$

where n indicates the refractive index of argon at the 632 nm probe beam wavelength of the dense gas at 200 bar [19]. Note that the probe beam is far detuned of the potassium D-lines, and the density of this gas is more than five orders of magnitude below that of the argon gas. The obtained deflection data are shown in Fig. 4 for an incident wavelength of 785 nm and an optical density of about 8.5 in the cell, corresponding to an optical absorption length $l_{abs}$ = 1.2 mm. The incident power of the cooling beam is about 1.5W, and the diameters of both laser beams are approximately 0.5 mm. The red line in Fig.4 shows a fit to the deflection data of a profile derived using a heat transport model [11, 18, 19, 24]. The black line is the corresponding temperature profile near the cell entrance, obtained by numerical integration of the data. We observe relative cooling by 120 K in the beam center near the entrance of the cell. This value is attributed to be limited by the thermal conductivity of the argon gas. Note that the cooling volume is not thermally isolated. The expected cooling power can be estimated from fluorescence spectra as recorded in Fig. 3, and at 785 nm incident laser wavelength is of order 30mW, corresponding to a cooling efficiency of 2%. For this estimation, we used $P_{cool} = P_{in} \cdot a(\upsilon) \cdot (\upsilon_{fl} - \upsilon)/\upsilon$, where $P_{in}$ denotes the incident optical laser power, $a(\upsilon)$ the frequency dependent absorption, $\upsilon$ the incident laser frequency and $\upsilon_{fl}$ the center frequency of the fluorescence, derived from Lorentzian fit curves to the data. On the other hand, the obtained value for the cooling power $P_{cool}$ is lower than the corresponding value obtained in the rubidium case [11], which we ascribe here to the faster drop-off of the far red wing of the line profile in the potassium-argon pair. The observed maximum relative temperature drop from the cooling (120 K for 785 nm incident laser wavelength) is about a factor of two above the result reported in our initial experiments demonstrating redistribution cooling [11]. To allow for a simple consideration of the scaling of the expected cooling in the limit of being determined by thermal conductivity, a simple estimate yields an expected temperature drop ΔT in the beam center for a cylindrical geometry in the stationary case

$$\Delta T(z) = \frac{P_{cool}}{2\pi\kappa} \cdot \alpha \cdot \ln\left(\frac{r_o}{r_i}\right) \cdot e^{-\alpha z},$$

where $r_i$ (≈ w) and $r_o$ denote inner and outer radii of integration, and for sake of simplicity one may set $r_o \approx l_{abs} = 1/\alpha$. The expected minimum temperature is mainly determined by the thermal conductivity $\kappa$ of the buffer gas, the cooling power $P_{cool}$, and the absorption length $l_{abs} = 1/\alpha$ (note that $l_{abs} \gg w$ is required to maintain a cylindrical geometry). Focussing the cooling beam more tightly as well as increasing the alkali vapour pressure, which leads to a reduced absorption length, can therefore lead to an enhanced decrease of the temperature. The use of heavier noble buffer gases, as krypton or xenon, would also be an option, since they have lower thermal conductivities than the presently used argon. The observed temperature drop of 120 K from the cooling is about a factor two above our results obtained in initial experiments on redistribution cooling (despite that the cooling efficiency is lower, see the above discussion, as attributed to different overlaps of quasimolecular wavefunctions of the rubidium-argon case with respect to the potassium-argon pair). We attribute the nevertheless improved temperature drop as being due to the shorter absorption length and the smaller size of the cooling laser beam used in the present experiment as compared to initial experiments (absorption length and beam diameter were factors of 4 and 2 respectively below the corresponding values given in [11]), which deposits the cooling power in a smaller volume within the gas and reduces the thermal conductivity. Cooling experiments were performed with several different wavelengths in the range from 780 nm to 795 nm. An incident wavelength of 785 nm (as described above) was observed to provide the most considerable temperature change for the used experimental parameters. For incident wavelengths of 780 nm, 790 nm and 795 nm, temperature changes of 103K, 75 K and 41 K were observed respectively.

## 5. Conclusion and Outlook

To conclude, we demonstrated redistribution laser cooling of a potassium-argon buffer gas mixture. The achieved temperature drop was 120 K at 200 bar buffer gas pressure in the beam center. Furthermore, the redistribution process in the ultrahigh buffer gas system has been resolved spectrally. Cooling to lower temperatures should be achievable in future experiments when the cooling laser is focused more tightly to a smaller beam diameter and using a smaller absorption length in the cell. As the Ti:Sapphire laser used in present experiments is tuneable in a wide spectral range, experiments on redistribution laser cooling of caesium atoms are also conceivable. Here, one could examine also the redistribution of fluorescence for the case of a broad fine structure splitting of the involved resonances. A detailed comparison between the alkali atoms for redistribution cooling in a high pressure buffer gas environment must also account for the corresponding spectral line shapes, which are determined by quasimolecular parameters. Future prospects of the described lasercooling can include cryocoolers and the study of homogeneous nucleation in saturated vapour [21-23].

We acknowledge support from the Deutsche Forschungsgemeinschaft within the focussed research unit FOR 557.

Fig 1: Cooling Principle. The black lines indicate the dependence how the energy levels of the potassium atoms change in presence of a noble gas perturber (see [12] for more realistic potential curves). Red detuned incident laser radiation can be absorbed when the distance between the alkali and the noble gas atom is small. Spontaneous decay occurs most probably at larger distances, so that the emitted photon has a frequency that is closer to the unperturbed resonance. The emitted fluorescence is blue shifted respectively to the incident radiation; therefore the atomic sample looses energy and is cooled.

Fig. 2: Experimental setup. a) Confocal setup spectrally resolved for measurements of atomic fluorescence in a high pressure buffer gas environment. The incident laser beam is spatially filtered by use of a confocal pinhole and then focused into the heated pressure cell. After spatial filtering, the collected atomic fluorescence is detected by an optical spectrum analyzer. b) Setup for thermal deflection spectroscopy. A cooling and an off-resonant probe beam are guided collinearly into the pressure cell, where temperature changes lead to a change of the refractive index and cause an angular deflection of a Helium-Neon-Laser beam. The deflected probe beam is detected subsequently by a position sensitive photodiode.

Fig. 3: Spectrum of the emitted fluorescence of potassium atoms at 200 bar argon buffer gas pressure. The red dashed and black solid lines show the spectrally resolved fluorescence for incident wavelengths of 780 nm (red detuned from the D-lines) and 765 nm (blue detuned) respectively. The position of the incident laser wavelength is indicated by vertical dotted lines for the two cases.

Fig 4: Measurement of gas temperature inside the cell. The black dots show measured data for the probe-beam angular deflection as a function of the lateral beam offset. The red line gives a fit to a profile derived using a heat transfer model. The solid black line shows the correspondingly derived temperature profile near the cell entrance. Relative cooling by an amount of 120 K is achieved in the beam center.

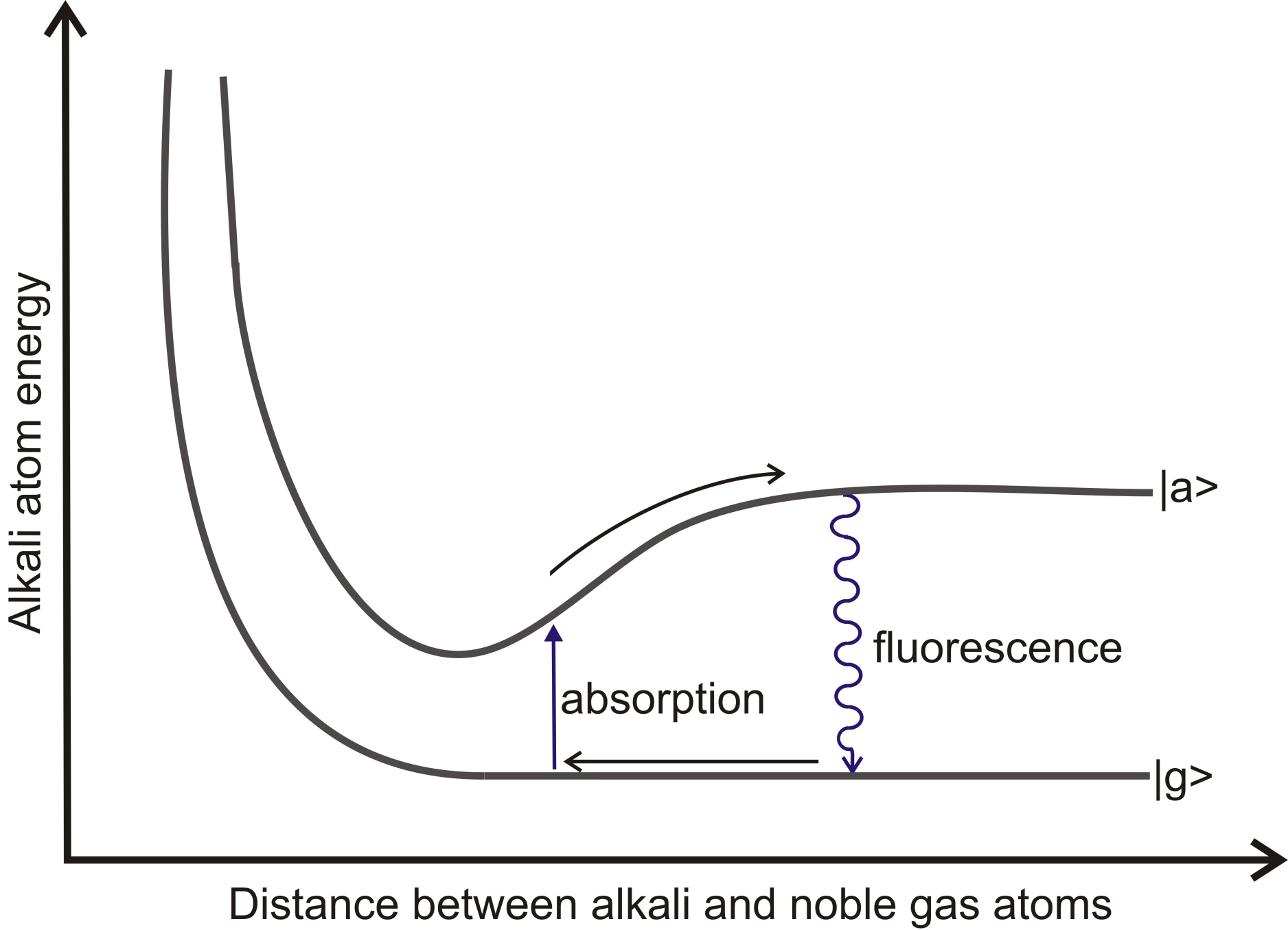

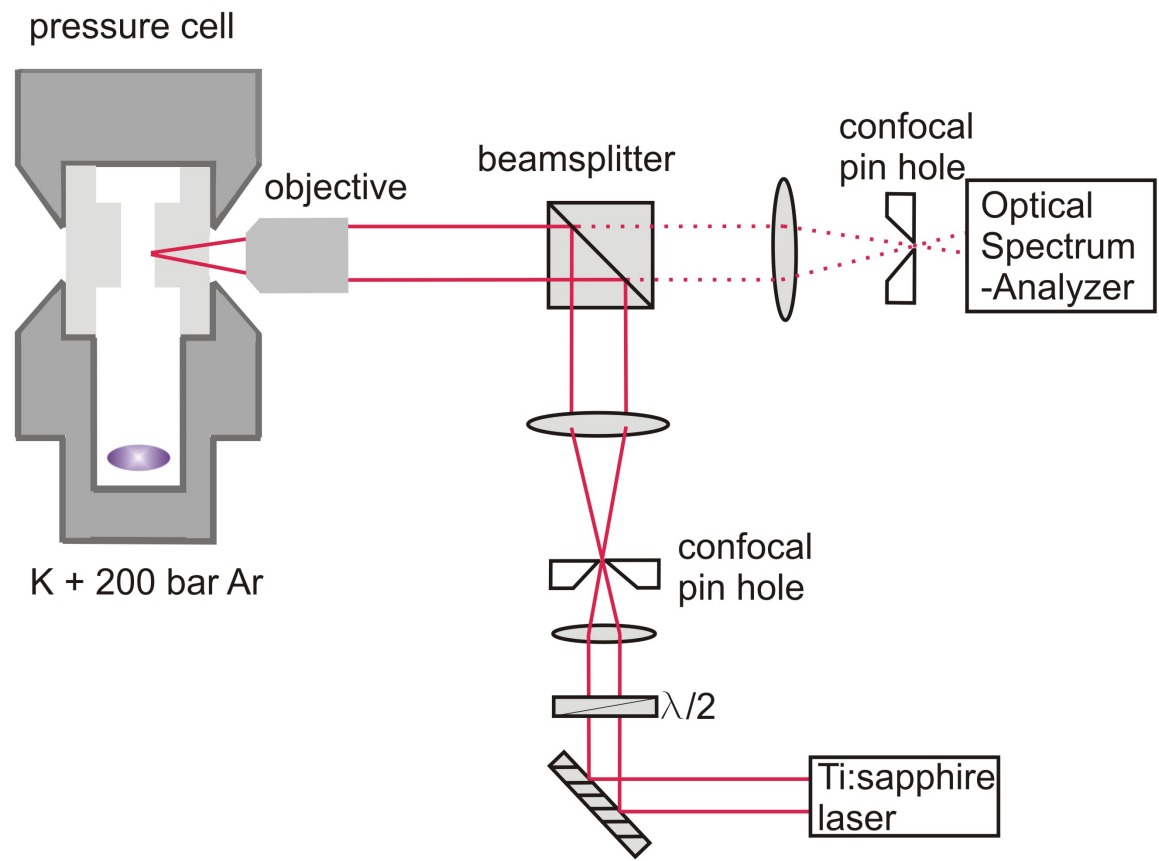 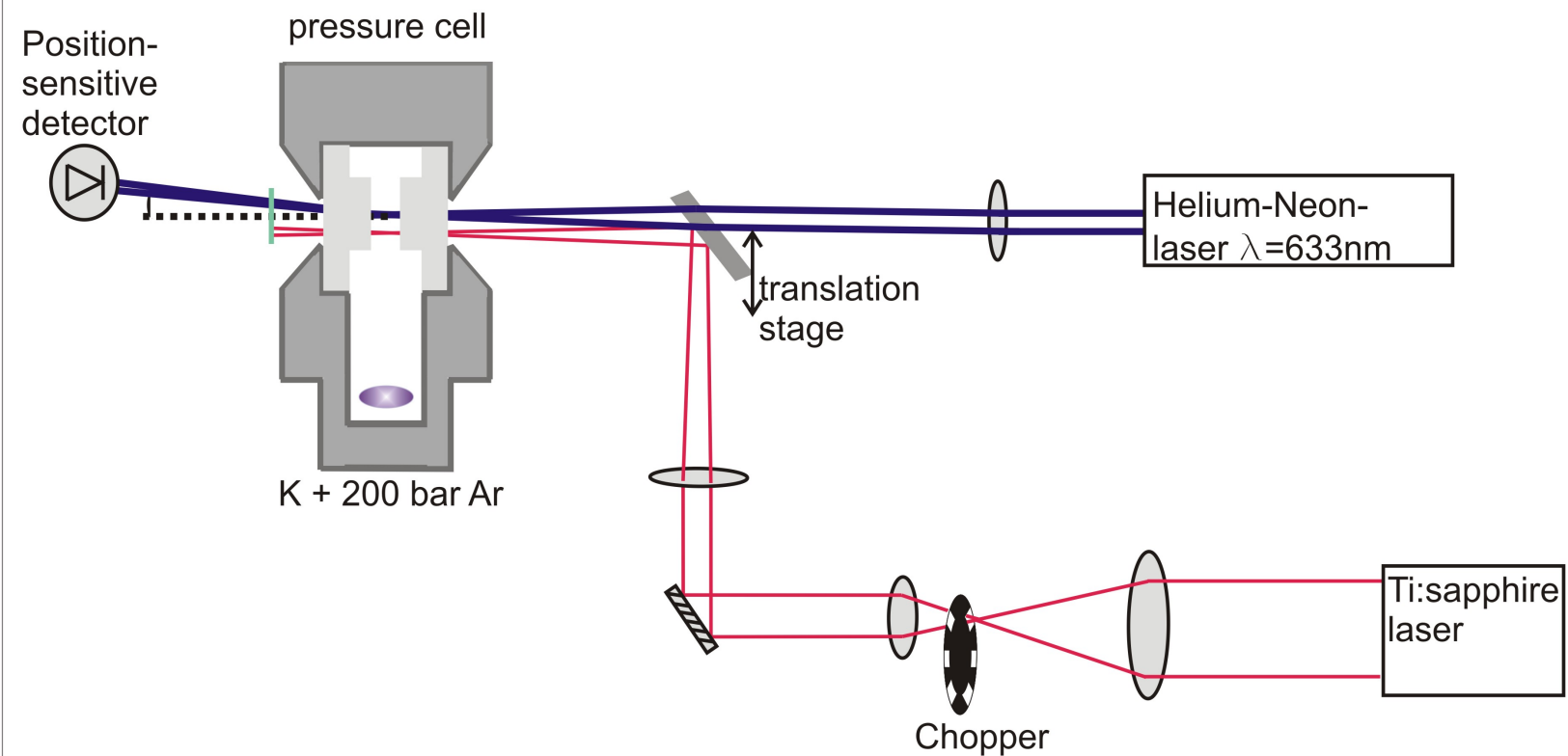

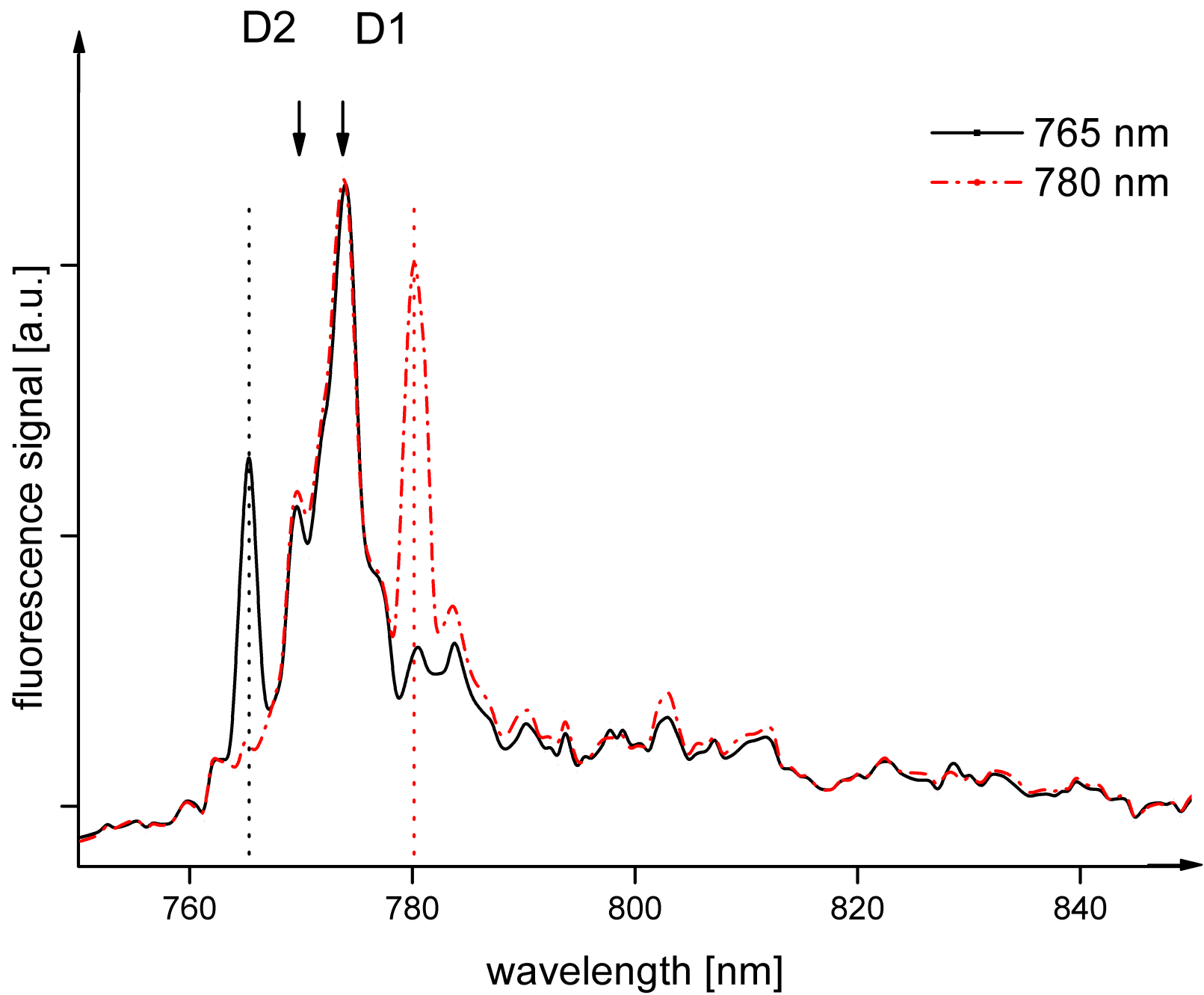

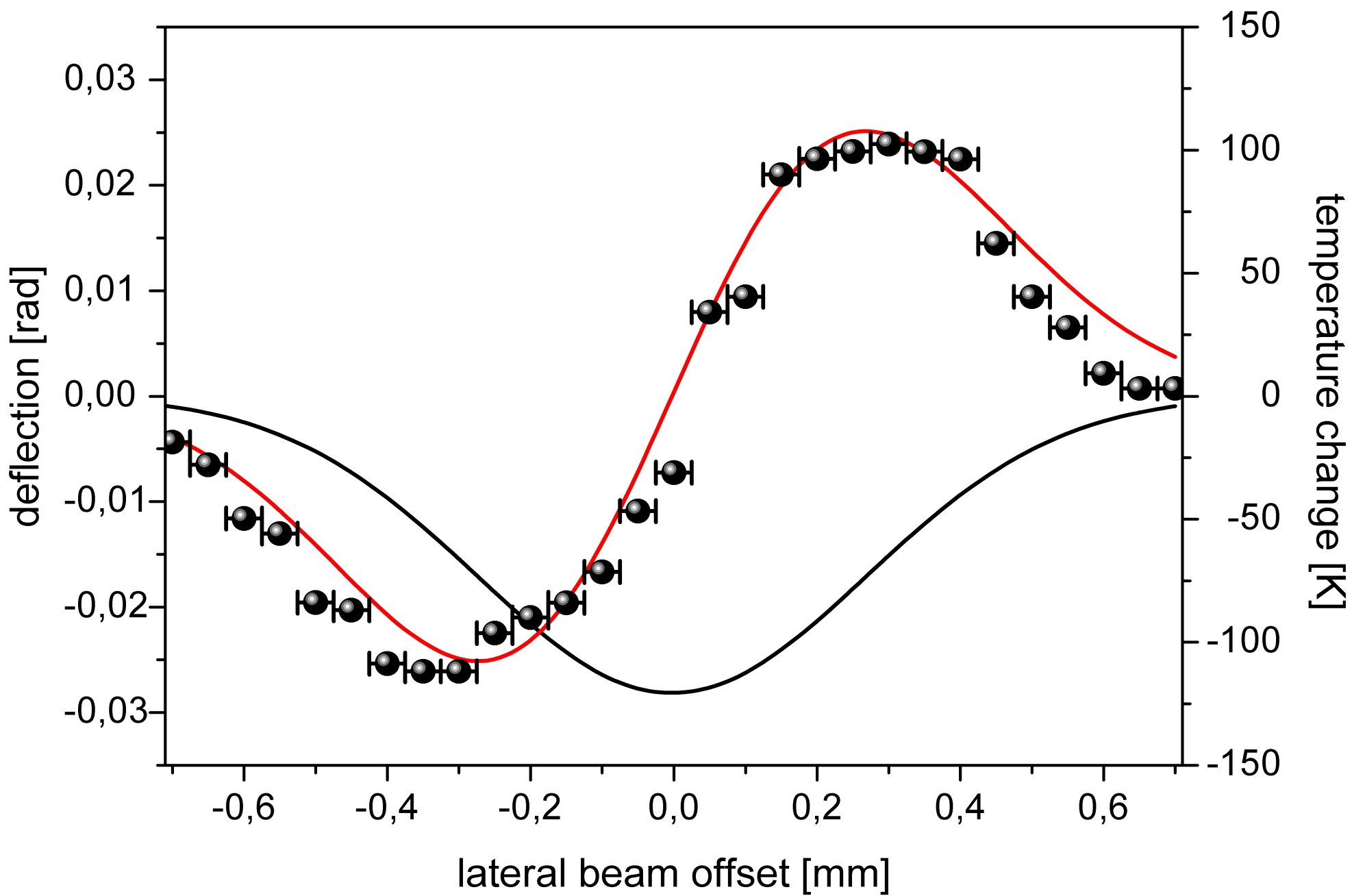